\def\rn{\noindent\parshape 2 0truecm 8.8truecm 0.3truecm 8.5truecm}
\def\nn#1 #2{#1, #2.}				
\def\nnn#1 #2 #3{#1, #2. #3.}			
\def\nnnn#1 #2 #3 #4{#1, #2. #3. #4.}		
\def\nnnnn#1 #2 #3 #4 #5{#1, #2. #3. #4. #5.}	
\def\dualand{, \&\hbox{ }}				
\def\multiand{, \&\hbox{ }}				

\def\rg#1;#2;#3;#4;#5;#6 {\par\rn#1 #2, {\it #3}, {\bf #4}, #5 (``#6'') \par}
\def\rf#1;#2;#3;#4;#5 {\par\rn#1 #2, {\it #3}, {\bf #4}, #5\par}
\def\rfbook#1;#2;#3;#4;#5 {{\frenchspacing\par\rn#1 #2, {\it #3} (#4: #5)\par}}
\def\rfproc#1;#2;#3;#4;#5;#6 {{\frenchspacing\par\rn#1 #2, in {\it #3}, ed. #4 (#5: #6)\par}}
\def\rfprocp#1;#2;#3;#4;#5;#6;#7 {{\frenchspacing\par\rn#1 #2, in {\it #3}, ed. #4 (#5: #6), p#7\par}}
\def\rfprep#1;#2;#3  {{\par\rn#1 #2, #3\par}}
\def\rfprepp#1;#2;#3 {{\par\rn#1 #2, #3\par}}

\def\Sec#1{Section~\ref{#1}}

\def\kg{{\rm kg}}

\def\R{{\bf R}}

\def\etal{{\frenchspacing\it et al.}}
\def\ie{{\frenchspacing\it i.e.}}
\def\eg{{\frenchspacing\it e.g.}}
\def\etc{{\frenchspacing\it etc.}}

\def\beq#1{\begin{equation}\label{#1}}
\def\eeq{\end{equation}}
\def\beqa#1{\begin{eqnarray}\label{#1}}
\def\eeqa{\end{eqnarray}}

\def\Sec#1{Section~\ref{#1}}

\def\spose#1{\hbox to 0pt{#1\hss}}
\def\simlt{\mathrel{\spose{\lower 3pt\hbox{$\mathchar"218$}}
     \raise 2.0pt\hbox{$\mathchar"13C$}}}
\def\simgt{\mathrel{\spose{\lower 3pt\hbox{$\mathchar"218$}}
     \raise 2.0pt\hbox{$\mathchar"13E$}}}
\def\simpropto{\mathrel{\spose{\lower 3pt\hbox{$\mathchar"218$}}
     \raise 2.0pt\hbox{$\propto$}}}

\def\ed{\end{document}}

\def\etal{{\frenchspacing\it et al.}}
\def\ie{{\frenchspacing\it i.e.}}
\def\eg{{\frenchspacing\it e.g.}}
\def\etc{{\frenchspacing\it etc.}}

\def\beq#1{$$}
\def\eeq{$$}

\def\fig#1{Figure~\ref{#1}}
\def\Fig#1{Figure~\ref{#1}}

\def\Sec#1{Section~\ref{#1}}

\documentclass[twocolumn,amsmath,nofootinbib]{revtex4} 
\usepackage{epsf} 
\begin{document}

\title{The Multiverse Hierarchy}

\author{Max Tegmark}

\address{Dept.~of Physics \& MIT Kavli Institute, Massachusetts Institute of Technology, Cambridge, MA 02139}
\date{February 20 2005, published in {\it ``Universe or Multiverse?"}, B.~Carr ed., Cambridge University Press, 2007)
}
\bigskip
\bigskip
\maketitle

\setcounter{secnumdepth}{2}

Parallel universes are now all the rage, cropping up in books, movies and 
even jokes: ``You passed your exam in many parallel universes --- but not this one."
However, they are as controversial as they are popular, and it is important to ask whether they are 
within the purview of science, or merely silly speculation. 
They are also a source of confusion, since many forget to distinguish between 
different types of parallel universes that have been proposed.

The farthest you can observe is the 
distance that light has been able to travel during the 14 billion years since the big-bang 
expansion began. The most distant visible objects are now about $4\times 10^{26}$ meters 
away\footnote{After emitting 
the light that is now reaching us, 
the most distant things we can see have receded because of the cosmic expansion,
and are now about about 40 billion light years away.
},
and a sphere of this radius defines our observable universe, also called our {\it Hubble volume}, our 
{\it horizon volume} or simply our universe. 
In this article, I survey physics theories involving parallel universes, which 
form a natural four-level hierarchy of multiverses (Figure 1) allowing progressively greater diversity.
\begin{itemize}
\itemsep-1mm
\item {\bf Level I:} A generic prediction of cosmological inflation is an infinite ``ergodic'' space,
which contains Hubble volumes realizing all initial conditions --- including
an identical copy of you about $10^{10^{29}}$m away.
\item {\bf Level II:} Given the {\it fundamental} laws of physics that physicists one day hope to capture with equations on a T-shirt, 
different regions of space can exhibit different
{\it effective} laws of physics (physical constants, dimensionality, particle content, \etc) corresponding
to different local minima in a landscape of possibilities. 
\item {\bf Level III:} In unitary quantum mechanics, other branches of the wavefunction 
add nothing qualitatively new, which is ironic given
that this level has historically been the most controversial.
\item {\bf Level IV:} Other mathematical structures give different fundamental equations of physics for that T-shirt.
\end{itemize}
The key question is therefore not whether there is a multiverse (since Level I is the rather uncontroversial 
cosmological concordance model), but rather how many levels it has.

Below we will discuss at length the issue of evidence and whether this is science or philosophy.
For now, the key point to remember is that
{\it parallel universes are not a theory, but a prediction of certain theories}.
For a theory to be falsifiable, we need not be able to observe and test all its predictions, merely at least one of them.
Consider the following analogy:
\begin{center}
\begin{tabular}{|l|l|}
\hline
General Relativity      &Black hole interiors\\
\hline
Inflation               &Level I parallel universes\\
\hline
Unitary quantum mechanics&Level III parallel universes\\
\hline
\end{tabular}
\end{center}
Because Einstein's theory of General Relativity has successfully predicted many things that we {\it can} observe, we also take seriously its predictions for things we
cannot observe, \eg, that space continues inside black hole event horizons and that (contrary to early misconceptions) nothing funny happens
right at the horizon.
Likewise, successful predictions of the theories of cosmological inflation and 
unitary\footnote{As described below, 
the mathematically simplest version of quantum mechanics is ``unitary'', lacking the 
controversial process known as wavefunction collapse.}
quantum mechanics have made some scientists take more seriously their other predictions, 
including various types of parallel universes.

Let us conclude with two cautionary remarks before delving into the details.
H\"ubris and lack of imagination have repeatedly caused us humans to underestimate the vastness of the physical world,
and dismissing things merely because we cannot observe them from our vantage point is reminiscent of the ostrich with its head in the sand.
Moreover, recent theoretical insights have indicated that Nature may be tricking us.
Einstein taught us that space is not merely a boring static void, but a dynamic entity that can stretch (the expanding universe), 
vibrate (gravitational waves) and curve (gravity). Searches for a unified theory also suggest that space can ``freeze'', 
transitioning between different phases in a landscape of possibilities just like water can be
solid, liquid or gas. In different phases, effective laws of physics (particles, symmetries, \etc.) could differ.
A fish never leaving the ocean might mistakenly conclude that the properties of water are universal, not realizing that there is also ice and steam.
We may be smarter than fish, but could be similarly fooled: cosmological inflation has the deceptive property of stretching a small patch of space in a particular phase 
so that it fills our entire observable universe, potentially tricking us into misinterpreting our local conditions for the universal laws that should go on that T-shirt.

\clearpage
\begin{figure}[tbp]
\vskip-1.5cm
\centerline{{\vbox{\hglue-0.5cm\epsfxsize=19.4cm\epsfbox{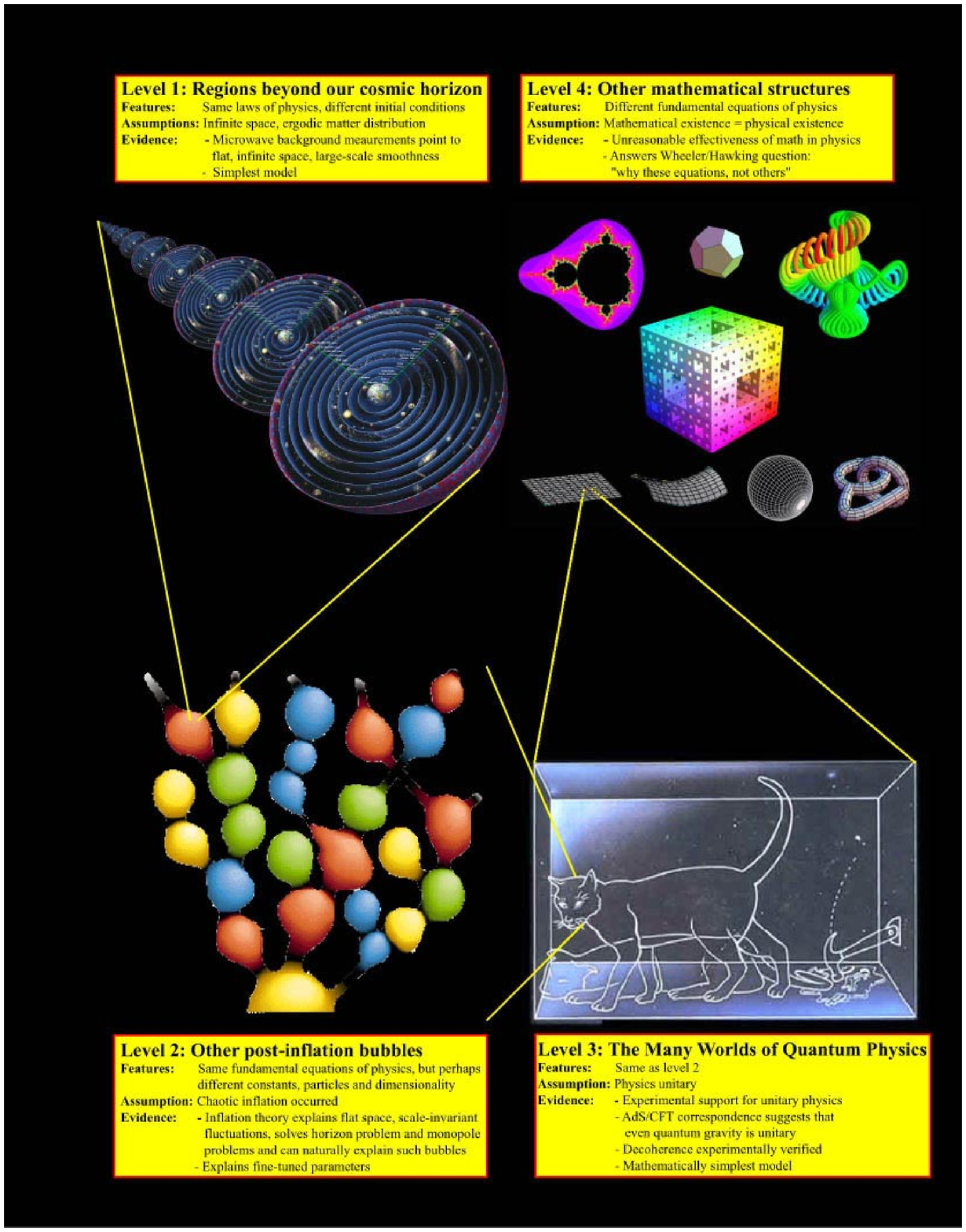}}}}
\label{ZoomFig}
\end{figure}
\setcounter{figure}{1}
\clearpage

\section{Level I: Regions beyond our cosmic horizon}

Let us return to your distant twin.
If space is infinite and the distribution of matter is sufficiently uniform on 
large scales, then even the most unlikely events must take place somewhere.
In particular, there are infinitely many other inhabited planets, including 
not just one but infinitely many with
people with the same appearance, name and memories as you.
Indeed, there are infinitely many other regions the size of our observable universe,
where every possible cosmic history is played out. This is the Level I multiverse.

\subsection{Evidence for Level I parallel universes}

Although the implications may seem crazy and counter-intuitive, this 
spatially infinite cosmological model is in fact the simplest and most popular one
on the market today. It is part of the cosmological concordance model, 
which agrees with all current observational evidence and is 
used as the basis for most calculations and simulations presented at cosmology conferences.
In contrast, alternatives such as a fractal universe, a closed universe and a multiply connected 
universe have been seriously challenged by observations.
Yet the Level I multiverse idea has been controversial (indeed, an assertion along these lines
was one of the heresies for which the Vatican had Giordano Bruno burned 
at the stake in 1600\footnote{Bruno's ideas have since been elaborated by, \eg,
Brundrit (1979), Garriga \& Vilenkin (2001b) and Ellis (2002), all of whom have
thus far avoided the stake.}), 
so let us review the status of the two assumptions
(infinite space and ``sufficiently uniform'' distribution). 
 
\begin{figure}[pbt]
\centerline{{\vbox{\hglue-5mm\epsfxsize=9.1cm\epsfbox{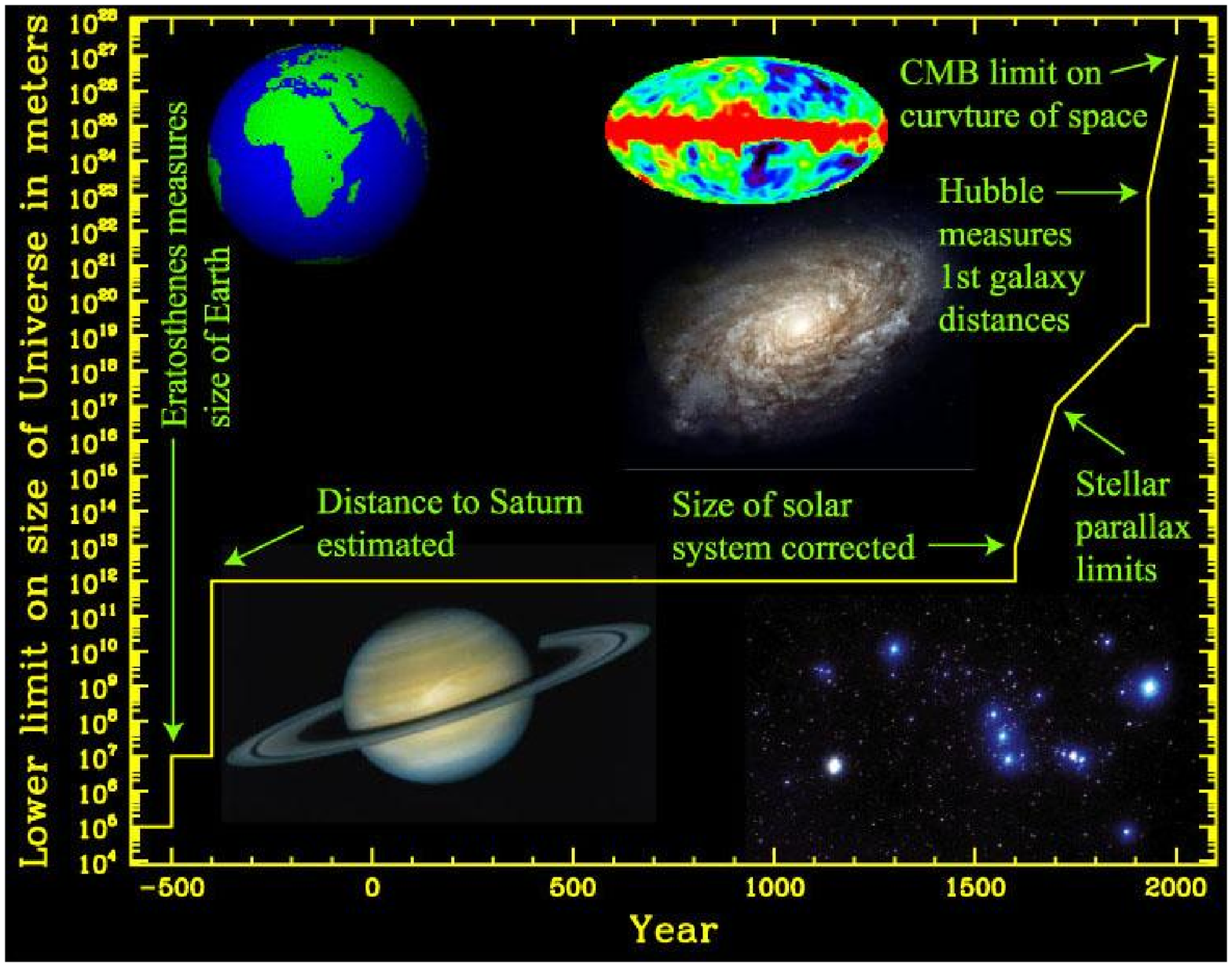}}}}
\vskip-2.5cm
\caption{Although an infinite universe has always been a possibility, the lower
limit on the size of our universe has kept growing.
}
\label{SizeFig}
\end{figure}

How large is space? Observationally, the lower bound has grown dramatically 
(\fig{SizeFig}) with no indication of an upper bound.
We all accept the existence of things that we cannot see but could see
if we moved or waited, like ships beyond the horizon. 
Objects beyond cosmic horizon have similar status, since the observable universe 
grows by a light-year every year as light from further away has time to 
reach us\footnote{If the cosmic expansion continues to accelerate (currently an open question),
the observable universe will eventually stop growing.
}. 
If anything, the Level I multiverse sounds trivially obvious. How could space not be
infinite? Is there a sign somewhere saying "Space Ends Here--Mind the Gap"? If so,
what lies beyond it? In fact, Einstein's theory of gravity calls this intuition into
question. Space could be finite if it has a convex curvature or an unusual topology
(that is, interconnectedness). A spherical, doughnut-shaped or pretzel-shaped
universe would have a limited volume and no edges. The cosmic microwave background
radiation allows sensitive tests of such scenarios. 
So far, however, the evidence is against them. Infinite models fit the
data, and strong limits have been placed on the alternatives 
(de Oliveira-Costa {\etal} 2003; Cornish {\etal} 2003).
In addition, 
a spatially infinite universe is a generic prediction of the cosmological theory of 
inflation (Garriga \& Vilenkin 2001b), so
the striking successes of inflation listed below therefore lend further support
to the idea that space is after all simple and infinite just as we learned in school.

Another loophole is that space is infinite but matter is confined to a finite region around us--the
historically popular "island universe" model. In a variant on this model, matter thins out on large scales in a
fractal pattern. In both cases, almost all universes in the Level I multiverse would be empty and dead. But
recent observations of the three-dimensional galaxy distribution and the microwave background have shown that
the arrangement of matter gives way to dull uniformity on large scales, with no coherent structures larger than
about 1024 meters. Assuming that this pattern continues, space beyond our observable universe teems with
galaxies, stars and planets.

\subsection{What are Level I parallel universes like?}
	\label{ErgodicitySec}

The physics description of the world is traditionally split into two parts:
initial conditions and laws of physics specifying how the initial conditions evolve.
Observers living in parallel universes at Level I observe the exact same laws of physics as we do,
but with different initial conditions than those in our Hubble volume.
The currently favored theory is that the initial conditions
(the densities and motions of different types of matter early on)
were created by quantum fluctuations during the inflation epoch (see section 3).
This quantum mechanism generates initial conditions that are for all practical 
purposes random, producing density fluctuations described by what mathematicians
call an ergodic random field.
{\it Ergodic} means that if you imagine 
generating an ensemble of universes, each with its own random initial conditions,
then the probability distribution of outcomes in a given volume
is identical to the distribution that you get by sampling different volumes in a single universe.
In other words, it means that everything that could in principle have happened here 
did in fact happen somewhere else.

Inflation in fact generates all possible initial conditions with non-zero probability, 
the most likely ones
being almost uniform with fluctuations at the $10^{-5}$ level that are amplified by
gravitational clustering to form galaxies, stars, planets and other structures.
This means both that pretty much all imaginable matter configurations occur in some Hubble 
volume far away, and also that we should expect our own Hubble volume to be a
fairly typical one --- at least typical among those that contain observers.
A crude estimate suggests that the closest identical copy of you is about $\sim10^{10^{29}}$m away. 
About $\sim 10^{10^{91}}$m away, there should be a sphere of radius 
100 light-years identical to the one centered here, so all perceptions that we have
during the next century will be identical to those of our counterparts over there.
About $\sim 10^{10^{115}}$m away, there should be an entire Hubble volume identical to 
ours.\footnote{This 
is an extremely conservative estimate, simply counting all possible quantum 
states that a Hubble volume can have that are no hotter than $10^8$K.  
$10^{115}$ is roughly the number of protons that the Pauli exclusion principle
would allow you to pack into a Hubble volume at this temperature
(our own Hubble volume contains only about $10^{80}$ protons).
Each of these $10^{115}$ slots can be either occupied or unoccupied, giving 
$N=2^{10^{115}}\sim 10^{10^{115}}$ possibilities, 
so the expected distance to the nearest identical Hubble volume is 
$N^{1/3}\sim 10^{10^{115}}$ Hubble radii $\sim 10^{10^{115}}$ meters.
Your nearest copy is likely to 
be much closer than $10^{10^{29}}$ meters, since the planet formation and evolutionary processes 
that have tipped the odds in your favor are at work everywhere. There are probably at 
least $10^{20}$ habitable planets in our own Hubble volume alone. 
}

This raises an interesting philosophical point that will come back and haunt us in \Sec{MeasureSec}: 
if there are indeed many copies of ``you'' with identical past lives and memories, 
you would not be able to compute your own future even if you had complete knowledge
of the entire state of the cosmos! 
The reason is that there is no way for you to determine 
which of these copies is ``you'' (they all feel that they are). Yet their lives will typically 
begin to differ eventually, so the best you can do is predict probabilities for what you will
experience from now on.  
This kills the traditional notion of determinism.

\subsection{How a multiverse theory can be tested and falsified}
\label{TestingSec}

Is a multiverse theory one of metaphysics rather than physics?
As emphasized by Karl Popper, the distinction between the two is whether the theory
is empirically testable and falsifiable.
Containing unobservable entities does clearly {\it not} per se make a theory non-testable.
For instance, a theory stating that there are 666 parallel universes, 
all of which are devoid of oxygen makes the testable prediction that we should observe no oxygen here, 
and is therefore ruled out by observation.

As a more serious example, the Level I multiverse framework is routinely used 
to rule out theories in modern cosmology, although this is rarely spelled out explicitly. 
For instance, cosmic microwave background (CMB) observations have recently 
shown that space has almost no curvature.
Hot and cold spots in CMB maps have a characteristic 
size that depends on the curvature of space, and the observed spots appear
too large to be consistent with the previously popular ``open universe'' model.
However, the average spot size randomly varies slightly from one Hubble 
volume to another, so it is important to be statistically rigorous.
When cosmologists say that the open universe model is ruled out at 
99.9\% confidence,
they really mean that if the open universe model were true,
then fewer than one out of every thousand Hubble volumes would show CMB spots as
large as those we observe --- therefore the entire model with all its infinitely many 
Hubble volumes is ruled out, even though we have of course only 
mapped the CMB in our own particular Hubble volume.

The lesson to learn from this example is that multiverse theories {\it can} be 
tested and falsified, but only if they predict 
what the ensemble of parallel universes is and specify  a probability distribution
(or more generally what mathematicians call a {\it measure})
over it. As we will see in \Sec{MeasureSec}, 
this measure problem can be quite serious and is still unsolved for
some multiverse theories.

\section{Level II: Other post-inflation bubbles}

If you felt that the Level I multiverse was large and hard to stomach,
try imagining an infinite set of distinct ones
(each symbolized by a bubble in 
Figure 1), some perhaps with different
dimensionality and different physical constants.
This is what is predicted by most 
currently popular models of inflation, and we will refer to
it as the Level II multiverse. 
These other domains are more than infinitely far away in the sense
that you would never get there even if you traveled at the speed of light forever.
The reason is that the space between our Level I multiverse and its neighbors
is still undergoing inflation, which keeps stretching it out 
and creating more volume faster than you can travel through it.
In contrast, you could travel to an arbitrarily distant Level I universe if you were patient 
and the cosmic expansion decelerates.\footnote{
Astronomical evidence suggests that the cosmic expansion is currently accelerating.
If this acceleration continues, then even the level I parallel universes will remain forever separate, 
with the intervening space stretching faster than light can travel through it.
The jury is still out, however, with popular models predicting that the universe will eventually 
stop accelerating and perhaps even recollapse.
}

\subsection{Evidence for Level II parallel universes}

Inflation is an extension of the big bang theory and ties up many of the loose ends of that theory, such as why
the universe is so big, so uniform and so flat. A rapid stretching of space long ago can explain all these and
other attributes in one fell swoop (see reviews by Linde 1994 and Guth \& Kaiser 2005).
Such stretching is predicted by a wide class of theories of elementary particles, and all available
evidence bears it out. Much of space is stretching and will continue doing so forever, 
but some regions of space stop stretching and form
distinct bubbles, like gas pockets in a loaf of rising bread. Infinitely many such bubbles emerge
(Figure 1, lower left, with time increasing upwards).
Each is an embryonic Level I multiverse: infinite in size\footnote{Surprisingly, it has been shown that inflation can produce 
an infinite Level I multiverse even in a bubble of finite spatial volume, thanks
to an effect whereby the spatial directions of spacetime curve towards the (infinite) 
time direction (Bucher \& Spergel 1999). 
} and filled with matter deposited by the energy field that drove inflation. 
Recent cosmological measurements have confirmed two key predictions of inflation: that space 
has negligible curvature and that the clumpiness in the cosmic matter distribution use to be approximately scale invariant.

\subsection{What are  Level II parallel universes like?}

The prevailing view is that the physics we observe today is
merely a low-energy limit of a much more general theory that 
manifests itself at extremely high temperatures. 
For example, this underlying fundamental theory
may be 10-dimensional, supersymmetric and involving a grand unification 
of the four fundamental forces of nature.
A common feature in such theories is that the potential energy of the
field(s) relevant to inflation has many different minima 
(sometimes called ``metastable vacuum states''), and ending up in different 
minima corresponds to different effective laws of physics for our low-energy world.
For instance, all but three spatial dimensions could be curled up (``compactified'') on a tiny scale, 
resulting in an effectively three-dimensional space like ours, or 
fewer could curl up leaving a 5-dimensional space.
Quantum fluctuations during inflation can therefore cause different post-inflation bubbles in
the Level II multiverse 
to end up with different effective laws of physics in different bubbles
--- say different dimensionality or different types of elementary particles, 
like two rather than three generations of quarks.

In addition to such discrete properties as dimensionality and particle content,
our universe is characterized by a set of dimensionless 
numbers known as {\it physical constants}. Examples include the electron/proton 
mass ratio $m_p/m_e\approx 1836$ and the cosmological constant, which appears to be about
$10^{-123}$ in so-called Planck units. There are models where also such 
non-integer parameters can vary from one one post-inflationary 
bubble to another.\footnote{
Although the fundamental equations of physics are the same throughout the Level II multiverse, the 
approximate effective equations governing the low-energy world that we observe will differ.
For instance, moving from a three-dimensional to a four-dimensional (non-compactified) space 
changes the observed gravitational force equation from an inverse square law to an inverse cube law.
Likewise, breaking the underlying symmetries of particle physics differently will change the lineup 
of elementary particles and the effective equations that describe them.
However, we will reserve the terms ``different equations'' and ``different laws of physics''
for the Level IV multiverse, where it is the fundamental rather than effective equations that change.
}
In summary, the Level II multiverse is likely to be more diverse than the Level I multiverse,
containing domains where not only the initial conditions differ,
but perhaps the dimensionality, the elementary particles and
the physical constants differ as well.

This is currently a very active research area.
The possibility of a string theory ``landscape'' (Bousso \& Polchinski 2000; Susskind 2003),
where the above-mentioned potential has perhaps $10^{500}$ different minima, may offer a specific realization of
the Level II multiverse which would in turn have four sub-levels of increasing diversity:
{\bf IId:} different ways in which space can be compactified, which can allow both different effective
dimensionality and different symmetries/elementary articles (corresponding to different 
topology of the curled up extra dimensions).
{\bf IIc:} different ``fluxes'' (generalized magnetic fields) that stabilize the extra dimensions (this sublevel is where the largest 
number of choices enter, perhaps $10^{500}$).
{\bf IIb:} once these two choices have been made, there may be a handful of different minima in the effective supergravity potential.
{\bf IIa:} the same minimum and effective laws of physics can be realized in a many different post-inflationary bubbles, each 
constituting a Level I multiverse.

Before moving on, let us briefly comment on a few closely related multiverse notions.
First of all, if one Level II multiverse can exist, eternally self-reproducing in a fractal pattern,
then there may well be infinitely many other Level II multiverses that are completely disconnected.
However, this variant appears to be untestable, since it would neither add any qualitatively different worlds 
nor alter the probability distribution for their properties.
All possible initial initial conditions and symmetry breakings
are already realized within each one.

An idea proposed by Tolman and Wheeler and recently elaborated by
Steinhardt \& Turok (2002) is that the (Level I) multiverse is
cyclic, going through an infinite series of Big Bangs.
If it exists, the ensemble of such incarnations would also form
a multiverse, arguably with a diversity similar to that of Level II.

An idea proposed by Smolin (1997) involves an ensemble similar in diversity to that of Level II, but 
mutating and sprouting new universes through black holes rather than during inflation.
This predicts a form of a natural selection favoring universes with maximal black hole production.

In braneworld scenarios, another 3-dimensional world could be quite
literally parallel to ours, merely offset in a higher dimension. However, it is unclear
whether such a world (``brane'') deserves be be called a parallel universe separate from our own, since 
we may be able to interact with it gravitationally much as we do with dark matter.

\begin{figure}[pbt]
\vskip-0.4cm
\centerline{{\vbox{\epsfxsize=8.8cm\epsfbox{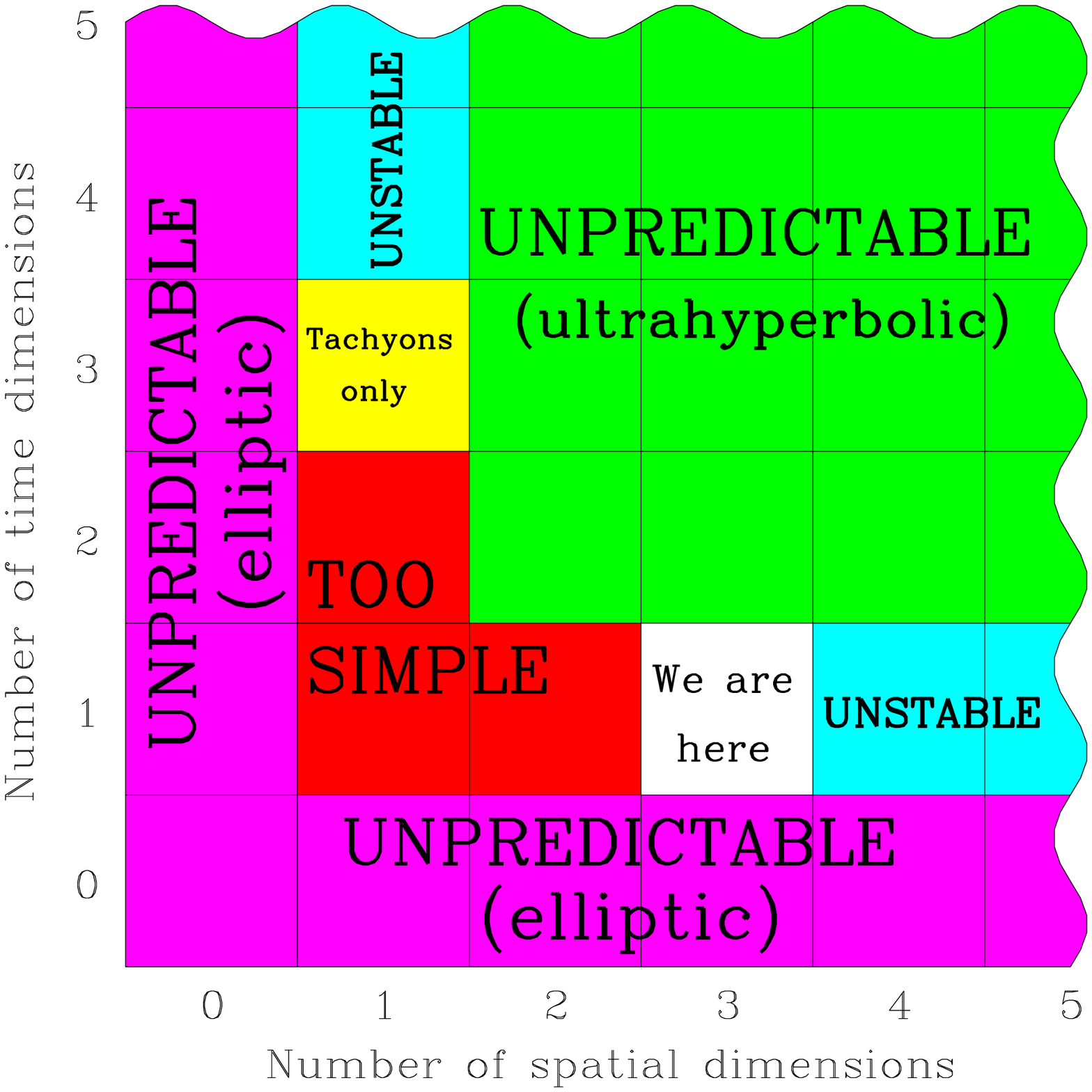}}}}
\smallskip
\caption{
\label{DimensionsFig} 
Why we should not be surprised to find ourselves living in 3+1-dimensional spacetime.
When the partial differential equations of nature are elliptic or
ultrahyperbolic, physics has no predictive power for an observer.
In the remaining (hyperbolic)
cases, $n>3$ admits no stable atoms and $n<3$ may lack sufficient complexity for observers
(no gravitational attraction, topological problems). From Tegmark (1997).
}
\vskip-0.1cm
\end{figure}

\subsection{Fine-tuning and selection effects}
\label{TuningSec}

Although we cannot interact with other Level II parallel universes, cosmologists can infer their
presence indirectly, because their existence can account for unexplained coincidences in our universe.
To give an analogy, suppose you check into a hotel, are assigned room 1967 and note that this is the
year you were born. What a coincidence, you say. After a moment of reflection, however, you conclude
that this is not so surprising after all. The hotel has hundreds of rooms, and you would not have been
having these thoughts in the first place if you had been assigned one with a number that meant nothing
to you. The lesson is that even if you knew nothing about hotels, you could infer the existence of other
hotel rooms to explain the coincidence. 

As a more pertinent example, consider the mass of the sun. The mass of a star determines its luminosity,
and using basic physics, one can compute that life as we know it on Earth is possible only if the sun's
mass falls into the narrow range between $1.6\times 10^{30}\kg$ and $2.4\times 10^{30}\kg$.
Otherwise Earth's climate
would be colder than that of present-day Mars or hotter than that of present-day Venus. The measured
solar mass is $M\sim 2.0\times 10^{30}\kg$. At first glance, this apparent coincidence of the habitable and
observed mass values appears to be a wild stroke of luck. Stellar masses run from $10^{29}$ 
to $10^{32}\kg$, so if the sun acquired its mass at random, it had only a small chance of falling into the
habitable range. But just as in the hotel example, one can explain this apparent coincidence by
postulating an ensemble (in this case, a number of planetary systems) and a selection effect (the fact
that we must find ourselves living on a habitable planet). Such observer-related selection effects are
referred to as ``anthropic'' (Carted 1973), and although the ``A-word'' is notorious for triggering controversy,
physicists broadly agree that these selection effects cannot be neglected when testing fundamental
theories. In this weak sense, the anthropic principle is not optional.

What applies to hotel rooms and planetary systems applies to parallel universes. Most, if not all, of the attributes set by
symmetry breaking appear to be fine-tuned. Changing their values by modest amounts would have resulted in a qualitatively
different universe--one in which we probably would not exist. If protons were $0.2\%$ heavier, they could decay into
neutrons, destabilizing atoms. If the electromagnetic force were $4\%$ weaker, there would be no hydrogen and no normal
stars. If the weak interaction were much weaker, hydrogen would not exist; if it were much stronger, supernovae would fail
to seed interstellar space with heavy elements. If the cosmological constant were much larger, the universe would have
blown itself apart before galaxies could form.
Indeed, most if not all the parameters affecting low-energy physics appear fine-tuned
at some level, in the sense that changing them by modest amounts results in 
a qualitatively different universe.

Although the degree of fine-tuning is still debated
(as exemplified in the rest of this book; see Barrow \& Tipler 1986, Tegmark 1998
\& Hogan (2000) for more technical reviews),
these examples suggest the existence of parallel universes with other
values of some physical constants. 
The existence of a Level II multiverse implies that physicists will never be able to determine the values of all
physical constants from first principles. Rather, they will merely compute probability distributions for what they 
should expect to find, taking selection effects into account. 
The result should be as generic as is consistent with our existence.

\begin{figure}[pbt]
\vskip-0.3cm
\centerline{{\vbox{\epsfxsize=8.8cm\epsfbox{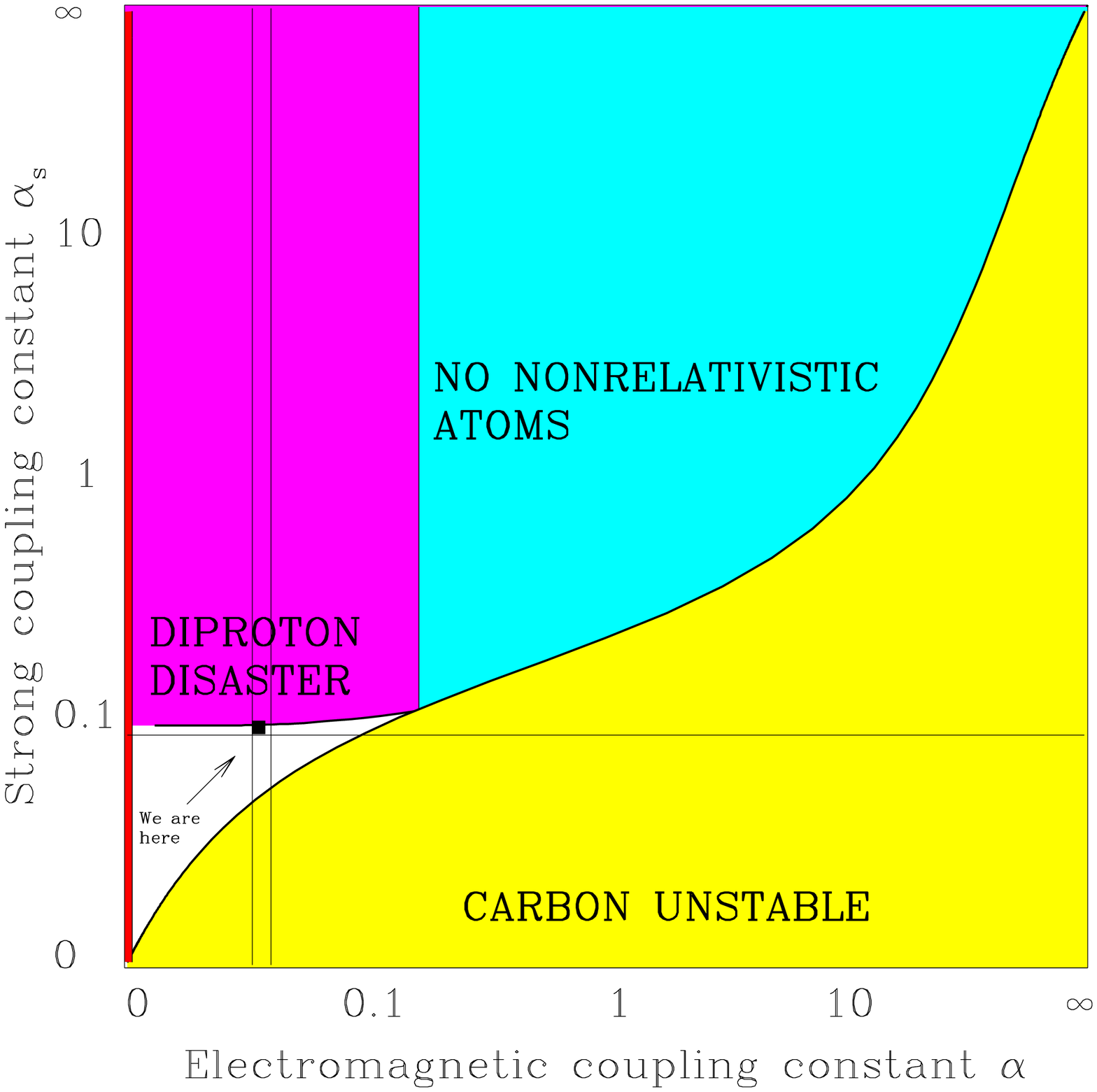}}}}
\smallskip
\caption{
Hints of fine-tuning for the parameters $\alpha$ and $\alpha_s$ which
determine the strengths of the 
electromagnetic force and the strong nuclear force, respectively
(from Tegmark 1997).
The observed values $(\alpha,\alpha_s)\approx(1/137,0.1)$
are indicated with a filled square. Grand unified theories
rule out everything except the narrow strip between the two vertical lines,
and deuterium becomes unstable below the horizontal line.
In the narrow shaded region to the very left,
electromagnetism is weaker than gravity and
therefore irrelevant.
\label{alphaalphasFig} 
}
\vskip-0.6cm
\end{figure}

\section{Level III: The many worlds of quantum physics}

There may be a third type of parallel worlds that are 
not far away but in a sense right here.
If the fundamental equations of physics are what mathematicians call {\it unitary}, 
as they so far appear to be, then the universe keeps branching into parallel universes 
as in the cartoon (\fig{CartoonFig}, bottom): whenever a quantum event appears to have a random outcome, 
all outcomes in fact occur, one in each branch.
This is the Level III multiverse. Although more debated and controversial than 
Level I and Level II, we will see that, surprisingly, 
this level adds no new types of universes.

\subsection{The quantum conundrum}

In the early 20th century the theory of quantum mechanics revolutionized physics by
explaining the atomic realm, which does not abide by the classical rules of Newtonian
mechanics. Despite the obvious successes of the theory, a heated debate rages about what
it really means. The theory specifies the state of the universe not in classical terms,
such as the positions and velocities of all particles, but in terms of a mathematical
object called a wave function. According to the Schr\"odinger equation, this state evolves
over time in a fashion that mathematicians term ``unitary'', meaning that the wave function
rotates in an abstract infinite-dimensional space called Hilbert space. Although quantum
mechanics is often described as inherently random and uncertain, the wave function
evolves in a deterministic way. There is nothing random or uncertain about it. 

\begin{figure}[pbt]
\centerline{{\vbox{\epsfxsize=8.7cm\epsfbox{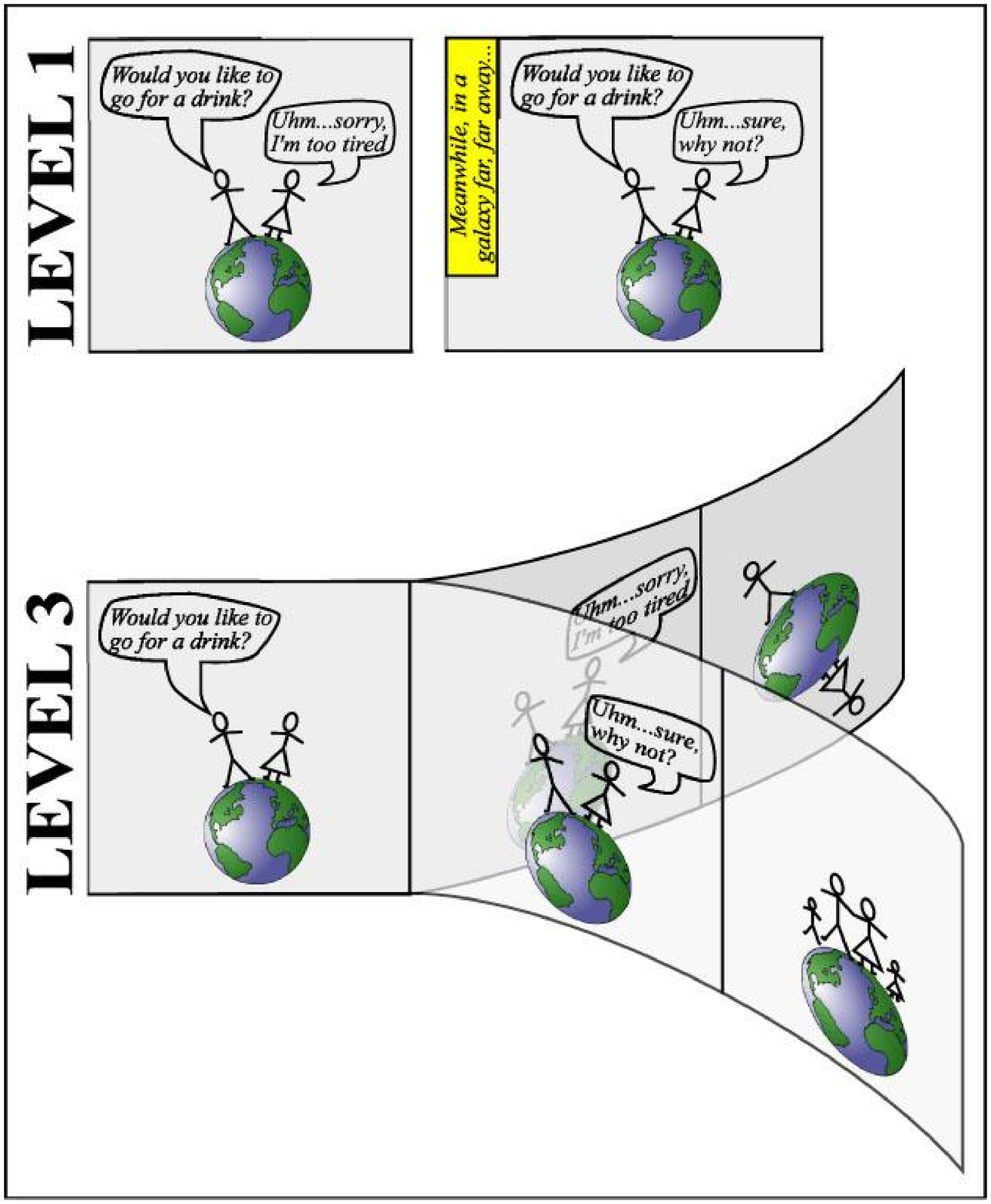}}}}
\smallskip
\caption{
Difference between Level I and Level III.
Whereas Level I parallel universes are far away in space, those of Level III
are even right here, with quantum events causing classical reality to split and diverge into 
parallel storylines. Yet Level III adds no new storylines beyond levels 1 or 2.
}
\label{CartoonFig}
\end{figure}

The sticky part is how to connect this wave function with what we observe. Many
legitimate wave functions correspond to counterintuitive situations, such as a cat being
dead and alive at the same time in a so-called superposition. In the 1920s physicists
explained away this weirdness by postulating that the wave function ``collapse'' into some
definite classical outcome whenever someone made an observation. This add-on had the
virtue of explaining observations, but it turned an elegant, unitary theory into a
kludgy, nonunitary one, since there was no equation specifying when or how this collapse occurred.
The intrinsic randomness commonly ascribed to quantum mechanics is the result of this postulate,
triggering Einstein's objection that ``God doesn't play dice''.

Over the years many physicists have abandoned this view in favor of one developed in 1957
by Princeton graduate student Hugh Everett III. He showed that the collapse postulate is
unnecessary. Unadulterated quantum theory does not, in fact, pose any contradictions.
Although it predicts that one classical reality gradually splits into superpositions of
many such realities, observers subjectively experience this splitting merely as a slight
randomness (\fig{CartoonFig}), with probabilities in exact agreement with those from the old collapse
postulate (de Witt 2003). This superposition of classical worlds is the Level III multiverse.

\subsection{What are Level III parallel universes like?}

Everett's many-worlds interpretation has been boggling minds inside and outside physics
for more than four decades. But the theory becomes easier to grasp when one distinguishes
between two ways of viewing a physical theory: the outside view of a physicist studying
its mathematical equations, like a bird surveying a landscape from high above it, and the
inside view of an observer living in the world described by the equations, like a frog
living in the landscape surveyed by the bird.\footnote{
Indeed, the standard mental picture of what the physical world is corresponds  
to a third intermediate viewpoint that could be termed the 
{\it consensus view}.
From your subjectively perceived frog perspective, the world turns upside down when you stand on your head
and disappears when you close your eyes, yet you subconsciously interpret your sensory inputs 
as though there is an external reality that is 
independent of your orientation, your location and your state of mind.
It is striking that although this third view involves both
censorship (like rejecting dreams),
interpolation (as between eye-blinks)
and extrapolation (say attributing existence to unseen cities)
of your inside view, independent observers nonetheless appear to share this 
consensus view.
Although the inside view looks black-and-white to a cat, 
iridescent to a bird seeing four primary colors,
and still more different to bee a seeing polarized light, a bat using sonar,
a blind person with keener touch and hearing, or the latest overpriced
robotic vacuum cleaner, all agree on whether the door is open.
The key current challenge in physics is deriving this semiclassical 
consensus view from the fundamental equations specifying the bird perspective.
In my opinion, this means that 
although understanding the detailed nature of human consciousness is an
important challenge in its own right, it is {\it not} necessary 
for a fundamental theory of physics.
}.

From the bird perspective, the Level III multiverse is simple. There is only one wave
function. It evolves smoothly and deterministically over time without any kind of
splitting or parallelism. The abstract quantum world described by this evolving wave
function contains within it a vast number of parallel classical story lines, continuously
splitting and merging, as well as a number of quantum phenomena that lack a classical
description. From their frog perspective, observers perceive only a tiny fraction of this
full reality. They can view their own Level I universe, but a process called
decoherence (Zeh 1970; Giulini {\etal} 1996) --- which mimics wave function collapse while preserving unitarity--prevents
them from seeing Level III parallel copies of themselves. 

Whenever observers are asked a question, make a snap decision and give an answer, quantum
effects in their brains lead to a superposition of outcomes, such as ``Continue reading
the article'' and ``Put down the article''. From the bird perspective, the act of making a
decision causes a person to split into multiple copies: one who keeps on reading and one
who doesn't. From their frog perspective, however, each of these alter egos is unaware of
the others and notices the branching merely as a slight randomness: a certain probability
of continuing to read or not. 

As strange as this may sound, the exact same situation occurs even in the Level I
multiverse. You have evidently decided to keep on reading the article, but one of your
alter egos in a distant galaxy put down the magazine after the first paragraph. The only
difference between Level I and Level III is where your doppelga{\"a}ngers reside. In Level I
they live elsewhere in good old three-dimensional space. In Level III they live on
another quantum branch in infinite-dimensional Hilbert space (\fig{CartoonFig}). 

\subsection{Level III parallel universes: evidence \& implications}

The existence of Level III depends on one crucial assumption: that the time evolution of the wave function is unitary. So
far experimenters have encountered no departures from unitarity. In the past few decades they have confirmed unitarity for
ever larger systems, including carbon 60 buckyball molecules and kilometer-long optical fibers. On the theoretical side,
the case for unitarity has been bolstered by the discovery of decoherence (see Tegmark \& 
Wheeler 2001 for a popular review). Some theorists who work on quantum gravity have
questioned unitarity; one concern is that evaporating black holes might destroy information, which would be a nonunitary
process. But a recent breakthrough in string theory known as AdS/CFT correspondence suggests that even quantum gravity is
unitary. If so, black holes do not destroy information but merely transmit it elsewhere.

If physics is unitary, then the standard picture of how quantum fluctuations operated early in the big bang must change.
These fluctuations did not generate initial conditions at random. Rather they generated a quantum superposition of all
possible initial conditions, which coexisted simultaneously. Decoherence then caused these initial conditions to behave
classically in separate quantum branches. Here is the crucial point: the distribution of outcomes on different quantum
branches in a given Hubble volume (Level III) is identical to the distribution of outcomes in different Hubble volumes
within a single quantum branch (Level I). This property of the quantum fluctuations is known in statistical mechanics as
ergodicity. 

The same reasoning applies to Level II. The process of symmetry breaking did not produce a unique outcome but rather a
superposition of all outcomes, which rapidly went their separate ways. So if physical constants, spacetime dimensionality
and so on can vary among parallel quantum branches at Level III, then they will also vary among parallel universes at
Level II. 

In other words, the Level III multiverse adds nothing new beyond Level I and Level II, just more indistinguishable copies
of the same universes--the same old story lines playing out again and again in other quantum branches. The passionate
debate about Everett's theory therefore seems to be ending in a grand anticlimax, with the discovery of less controversial
multiverses (Levels I and II) that are equally large. 

Needless to say, the implications are profound, and physicists are only beginning to explore them. For instance, consider
the ramifications of the answer to a long-standing question: Does the number of universes exponentially increase over
time? The surprising answer is no. From the bird perspective, there is of course only one quantum universe. 

From the frog perspective, what matters is the number of universes that are distinguishable at a given instant--that is,
the number of noticeably different Hubble volumes. Imagine moving planets to random new locations, imagine having married
someone else, and so on. At the quantum level, there are 10 to the $10^{118}$ universes with temperatures 
below $10^8$ kelvins.
That is a vast number, but a finite one. 

From the frog perspective, the evolution of the wave function corresponds to a never-ending sliding from one of these 
$10$ to the $10^{118}$ states to another. 
Now you are in universe A, the one in which you are reading this sentence. Now you are in
universe B, the one in which you are reading this other sentence. Put differently, universe B has an observer identical to
one in universe A, except with an extra instant of memories. All possible states exist at every instant, so the passage of
time may be in the eye of the beholder --- an idea explored in Greg Egan's 1994 science-fiction novel Permutation City and
developed by physicist David Deutsch of the University of Oxford, independent physicist Julian Barbour, and others. The
multiverse framework may thus prove essential to understanding the nature of time.

\begin{figure}[pbt]
\hglue-1.7cm\centerline{{\vbox{\epsfxsize=12.0cm\epsfbox{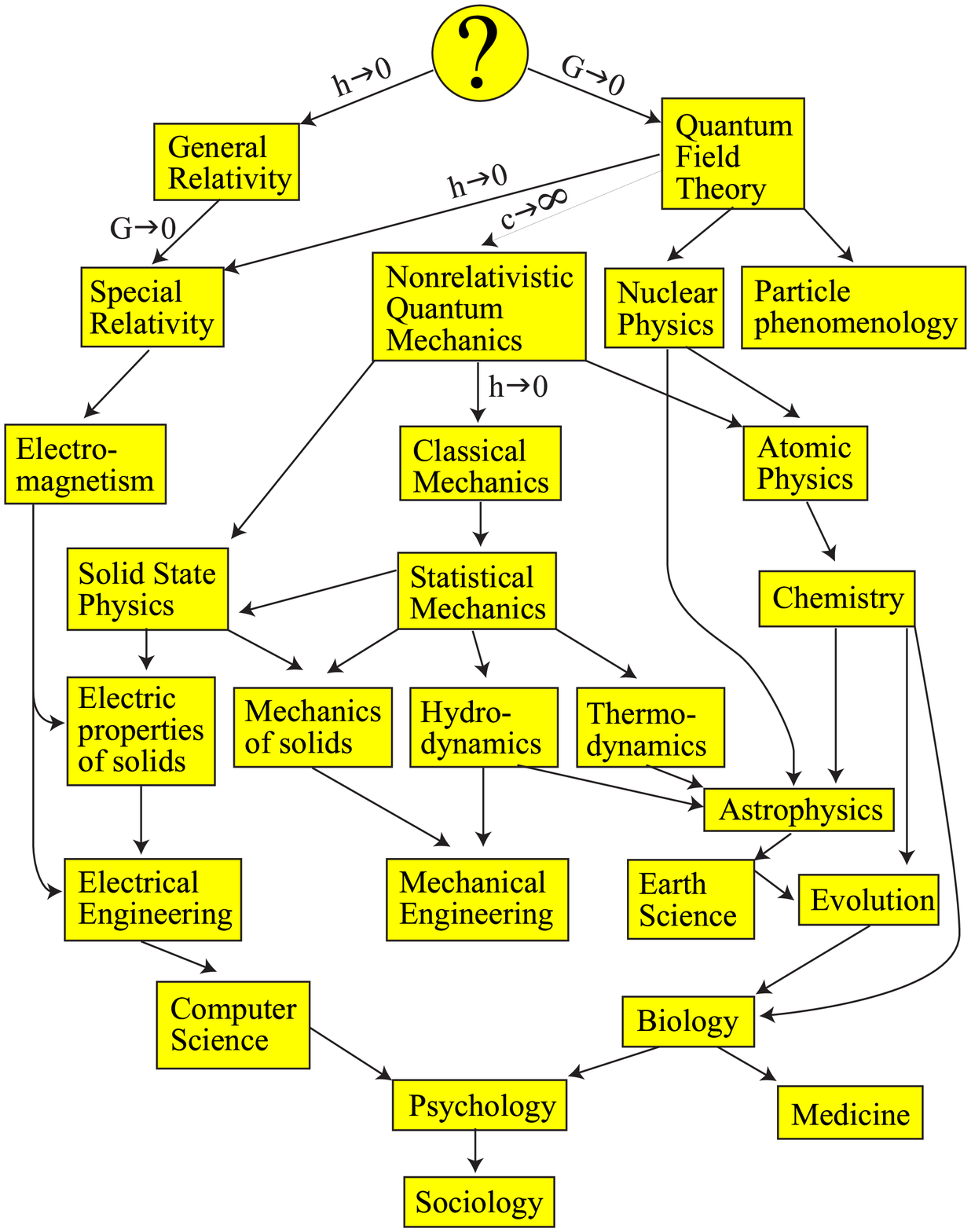}}}}
\smallskip
\caption{
Theories can be crudely organized into a family tree where
each might, at least in principle, be derivable
from more fundamental ones above it.
For example, classical mechanics can be obtained from 
special relativity in the approximation that the speed of light 
$c$ is infinite. Most of the arrows are less well understood.
All these theories have two components: mathematical equations and words
that explain how they are connected to what we observe. 
At each level in the hierarchy of theories, new words (e.g.,
protons, atoms, cells, organisms, cultures) are introduced because they
are convenient, capturing the essence of what is going on without
recourse to the more fundamental theory above it.  It is important to
remember, however, that it is we humans who introduce these concepts and
the words for them: in principle, everything could have been derived from
the fundamental theory at the top of the tree, although such an extreme
reductionist approach would of course be useless in practice. Crudely
speaking, the ratio of equations to words decreases as we move down the
tree, dropping near zero for highly applied fields such as medicine and
sociology. In contrast, theories near the top are highly mathematical,
and physicists are still struggling to understand the concepts, if any,
in terms of which we can understand them.
The Holy Grail of physics is to find what is jocularly referred to as a
``Theory of Everything'', or TOE, from which all else can be derived. If
such a theory exists at all, it should replace the big question mark at
the top of the theory tree. Everybody knows that something is missing
here, since we lack a consistent theory unifying gravity with quantum
mechanics. 
}
\label{TheoryTreeFig}
\end{figure}

\subsection{Two world views}

The debate over how classical mechanics emerges from quantum mechanics
continues, and the decoherence discovery has shown that there is 
a lot more to it than just letting Planck's constant $\hbar$ shrink to zero.
Yet as \fig{TheoryTreeFig}  illustrates, this is just a small piece of a larger puzzle.
Indeed, the endless debate over the interpretation of quantum mechanics 
--- and even the broader issue of parallel universes --- is in a sense the
tip of an iceberg.
In the Sci-Fi spoof ``Hitchhiker's Guide to the Galaxy'', the answer 
is discovered to be ``42'', and the hard part is finding the real question.
Questions about parallel universes may seem to be just about as deep as queries about 
reality can get. Yet there is a still deeper underlying question:
there are two tenable but diametrically opposed
paradigms regarding physical reality and the status of mathematics, a dichotomy that  
arguably goes as far back as Plato and Aristotle, and the question is which one is correct.
\begin{itemize}
\item {\bf ARISTOTELIAN PARADIGM:} The subjectively perceived frog perspective is physically real, 
and the bird perspective and all its mathematical language is 
merely a useful approximation.
\item {\bf PLATONIC PARADIGM:} The bird perspective (the mathematical structure) is physically 
real, and the frog perspective and all the human language we use
to describe it is merely a useful approximation for describing
our subjective perceptions.
\end{itemize}
What is more basic --- the frog perspective or the bird perspective?
What is more basic --- human language or mathematical language?
Your answer will determine how you feel about parallel universes.
If you prefer the Platonic paradigm, you should find multiverses natural, 
since our feeling that say the Level III multiverse is ``weird'' merely reflects that
the frog and bird perspectives are extremely different. We break the symmetry by calling the 
latter weird because we were all indoctrinated with the Aristotelian 
paradigm as children, long before we even heard of mathematics - the Platonic view is an acquired taste!

In the second (Platonic) case, all of physics is ultimately a mathematics problem, since 
an infinitely intelligent mathematician given the fundamental equations of
the cosmos could in principle
{\it compute} the frog perspective, {\ie}, 
compute what self-aware observers the universe would contain, 
what they would perceive, and what language they would 
invent to describe their perceptions to one another. 
In other words, there is a  ``Theory of Everything" (TOE) at the top of the tree in \fig{TheoryTreeFig} 
whose axioms are purely mathematical, since postulates in English 
regarding interpretation would be derivable and thus redundant.
In the Aristotelian paradigm, on the other hand, there can never be a 
TOE, since one is ultimately just explaining
certain verbal statements by other verbal statements ---
this is known as the infinite regress problem (Nozick 1981).

\section{Level IV: Other mathematical structures}

Suppose you buy the Platonist paradigm and believe that
there really is a TOE at the top of \fig{TheoryTreeFig}  --- and that we simply have not found
the correct equations yet.
Then an embarrassing question remains, as emphasized by John Archibald
Wheeler: {\it  Why these particular equations, not others?}
Let us now explore the idea of mathematical democracy, whereby
universes governed by other equations are equally real.
This is the Level IV multiverse.
First we need to digest two other ideas, however:
the concept of a mathematical structure, and the notion that the physical world may be one.

\subsection{What is a mathematical structure?}

Many of us think of mathematics as a bag of tricks 
that we learned in school for manipulating numbers.
Yet most 
mathematicians have a very different view of their field.
They study more abstract objects such as functions, sets, spaces and operators
and try to prove theorems about the relations between them.
Indeed, some modern mathematics papers are so abstract that the only numbers you will
find in them are the page numbers!
What does a dodecahedron have in common with a set of complex numbers?
Despite the plethora of mathematical structures with intimidating names like 
orbifolds and Killing fields, a striking underlying unity has emerged in the last century:
{\it all} mathematical structures are just special cases of one and the same 
thing: so-called formal systems.
A formal system consists of abstract symbols and rules for manipulating them, 
specifying how new strings of symbols referred to as theorems can be derived from
given ones referred to as axioms.
This historical development represented a form of deconstructionism, since it
stripped away all meaning and interpretation that had traditionally been given to
mathematical structures and distilled out only the 
abstract relations capturing their very essence.
As a result, computers can now prove theorems about geometry
without having any physical intuition whatsoever about what space is like.

\begin{figure}[pbt]
\centerline{{\vbox{\epsfxsize=8.5cm\epsfbox{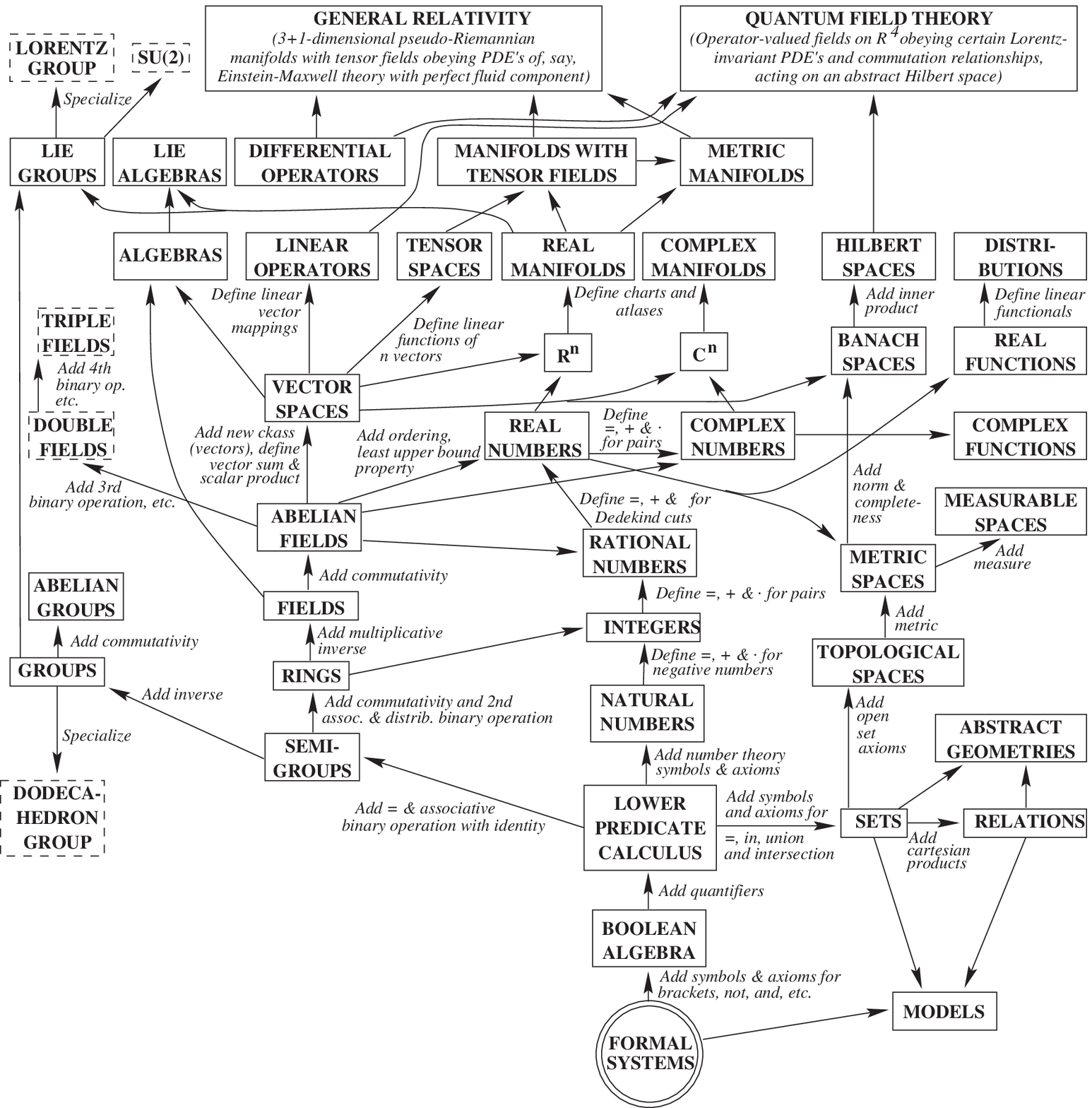}}}}
\smallskip
\caption{
\label{MathTreeFig}
Relationships between various basic mathematical structures (Tegmark 1998).
The arrows generally indicate addition of new symbols and/or axioms.
Arrows that meet indicate the combination of structures
--- for instance,
an algebra is a vector space that is also a ring,
and a Lie group is a group that is also a manifold.
The full tree is probably infinite in extent --- the figure shows
merely a small sample near the bottom.
}
\end{figure}

\Fig{MathTreeFig} shows some of the most basic mathematical structures and
their interrelations. Although this family tree probably extends indefinitely, it illustrates
that there is nothing fuzzy about mathematical structures. 
They are ``out there'' in the sense that mathematicians discover them rather than create them,
and that contemplative alien civilizations would find the same structures
(a theorem is true regardless of whether it is proven by a human, a computer or an alien).

\subsection{The possibility that the physical world is a mathematical structure}

Let us now digest the idea that physical world (specifically, the Level III multiverse)  
{\it is} a mathematical structure.
Although traditionally taken for granted by many theoretical physicists, this is a deep 
and far-reaching notion. It means that mathematical equations describe not merely some
limited aspects of the physical world, but {\it all} aspects of it.
It means that there is some mathematical structure that is what mathematicians call
{\it isomorphic} (and hence equivalent) to our physical world, with each physical
entity having a unique counterpart in the mathematical structure and vice versa.
Let us consider some examples.

A century ago, when classical physics still reigned supreme, 
many scientists believed that physical space was isomorphic 
to the mathematical structure known as $\R^3$:
three-dimensional Euclidean space. 
Moreover, some thought that all forms of matter in the universe
corresponded to various classical {\it fields}: the electric field, the magnetic field and 
perhaps a few undiscovered ones, mathematically corresponding to functions on $\R^3$
(a handful of numbers at each point in space).
In this view (later proven incorrect), dense clumps of matter like atoms were simply 
regions in space where some fields were strong (where some numbers were large).
These fields evolved deterministically over time according to some partial differential equations,
and observers perceived this as things moving around and events taking place.
Could, then, fields in three-dimensional space be the mathematical structure corresponding
to the universe? No,  since a mathematical structure cannot change --- it is an abstract,
immutable entity existing outside of space and time. 
Our familiar frog perspective of a three-dimensional space where events unfold
is equivalent, from the bird perspective, to a four-dimensional spacetime where all of
history is contained, 
so the mathematical 
structure would be fields in four-dimensional space.
In other words, if history were a movie, the mathematical structure would 
not correspond to a single frame of it, but to the entire videotape.

Given a mathematical structure, we will say that it has {\it physical existence}
if any self-aware substructure (SAS) within it subjectively, from its frog perspective,
perceives itself as living in a physically real world.
What would, mathematically, such an SAS be like?
In the classical physics example above, an SAS such as you would be 
a tube through spacetime, a thick version of what Einstein referred to as a  world-line.
The location of the tube would specify your position in space at different times.
Within the tube, the fields would exhibit certain complex behavior, corresponding
to storing and processing information about the field-values in the surroundings,
and at each position along the tube, these processes would give rise to 
the familiar but mysterious sensation of self-awareness.
From its frog perspective, the SAS would perceive this one-dimensional 
string of perceptions along the tube as passage of time.

Although our example illustrates the idea of how our physical world can {\it be} a mathematical
structure, this particular mathematical structure (fields in four-dimensional space) is now known
to be the wrong one. After realizing that spacetime could be curved, Einstein doggedly searched
for a so-called unified field theory where the universe was what mathematicians call 
a 3+1-dimensional pseudo-Riemannian manifold with tensor fields (top center in \fig{MathTreeFig}), but this failed to
account for the observed behavior of atoms. According to quantum field theory, the
modern synthesis of special relativity theory and quantum theory, 
the universe (in this case the Level III multiverse) 
is a mathematical structure known as an algebra of operator-valued 
fields (top right in \fig{MathTreeFig}). Here the question of what constitutes an
SAS is more subtle (Tegmark 2000).
However, this fails to describe black hole evaporation, the first instance of the Big Bang 
and other quantum gravity phenomena, so 
the true mathematical structure isomorphic to our universe, if it exists, has not yet been found.

\subsection{Mathematical democracy}

Now suppose that our physical world really is a mathematical structure, 
and that you are an SAS within it. This means that in
the Mathematics tree of \fig{MathTreeFig}, one of the boxes is our universe.
(The full tree is probably infinite in extent, so our particular box is not one of the few boxes 
from the bottom of the tree that are shown.)
In other words, this particular mathematical structure enjoys not only mathematical
existence, but physical existence as well.
What about all the other boxes in the tree? Do they too enjoy physical existence? If not, there would be a fundamental, unexplained ontological asymmetry built into the
very heart of reality, splitting mathematical structures into two classes: 
those with and without physical existence. 
As a way out of this philosophical conundrum, I have suggested (Tegmark 1998)
that complete mathematical democracy holds: that mathematical existence and 
physical existence are equivalent, so that  {\it all} mathematical structures exist 
physically as well.
This is the Level IV multiverse. It can be 
viewed as a form of radical Platonism,
asserting that the mathematical structures
in Plato's {\it realm of ideas}, the {\it Mindscape} of Rucker (1982), exist 
``out there'' in a physical sense (Davies 1993),
casting the so-called modal realism theory of  
David Lewis (1986) in mathematical terms
akin to what Barrow (1991; 1992) refers to as ``$\pi$ in the sky''.
If this theory is correct, then since it has no free parameters,
all properties of all parallel universes
(including the subjective perceptions of SASs in them) could in principle be derived by
an infinitely intelligent mathematician.

\subsection{Evidence for a Level IV multiverse}

We have described the four levels of parallel universes in order of increasing
speculativeness, so why should we believe
in Level IV? Logically, it rests on two separate assumptions:
\begin{itemize}
\item {\bf Assumption 1:} That the physical world (specifically our level III multiverse) is a mathematical structure
\item {\bf Assumption 2:} Mathematical democracy: that all mathematical structures  exist ``out there'' in the same sense
\end{itemize}

In a famous essay, Wigner (1967) argued that 
``the enormous usefulness of mathematics in the
natural sciences is something bordering on the mysterious", and that
``there is no rational explanation for it".
This argument can be taken as support for assumption 1: here
the utility of mathematics for describing the physical
world is a natural consequence of the fact that the latter {\it is}
a mathematical structure, and we are simply uncovering this 
bit by bit.
The various approximations that constitute our current physics theories
are successful because simple mathematical
structures can provide good approximations
of how a SAS will perceive more complex mathematical structures.
In other words, our successful theories are
not mathematics approximating physics,
but mathematics approximating mathematics.
Wigner's observation is unlikely to be based on fluke coincidences,
since far more mathematical regularity in nature has been discovered in the 
decades since he made it, including the standard model of particle physics.

A second argument supporting assumption 1 is 
that abstract mathematics is so general that 
{\it any} TOE that is definable in purely formal terms
(independent of vague human terminology)
is also a mathematical structure.
For instance,
a TOE involving a set of different types of entities
(denoted by words, say) and
relations between them (denoted by additional words)
is nothing but what mathematicians call
a set-theoretical model, and one can generally find a
formal system that it is a model of.

This argument also makes assumption 2 more appealing,
since it implies that {\it any} conceivable parallel universe theory can be described 
at Level IV. The Level IV multiverse, termed the ``ultimate Ensemble theory'' in 
Tegmark (1997) since it subsumes
all other ensembles, therefore brings closure to the 
hierarchy of multiverses, and there cannot be say a Level V.
Considering an ensemble of mathematical structures does not add anything new, since
this is still just another mathematical structure.
What about the frequently discussed notion that the universe is a computer simulation?
This idea occurs frequently in science fiction and has been substantially
elaborated (\eg, Schmidthuber 1997; Wolfram 2002).
The information content (memory state) of a digital computer is a string of bits, say 
``$1001011100111001...$'' of great but finite length, equivalent to some large 
but finite integer $n$ written in binary. The information processing of a computer is
a deterministic rule for changing each memory state into another (applied over and over again),
so mathematically, it is simply a function $f$ mapping the integers onto themselves
that gets iterated: $n\mapsto f(n)\mapsto f(f(n))\mapsto...$.
In other words, even the most sophisticated computer simulation is
just yet another special case of a mathematical structure,
and is already included in the Level IV multiverse.
(Incidentally, iterating continuous functions rather than integer-valued ones
can give rise to fractals.)

A second argument for assumption 2 is that if two entities are isomorphic, then there is 
no meaningful sense in which they are not one and the same (Cohen 2003). 
This implies assumption 2 when the entities in question are a physical universe and
a mathematical structure describing it, respectively.
To avoid this conclusion that mathematical and physical existence are equivalent,
one would need to argue that our universe is somehow made of stuff perfectly described
by a mathematical structure, but which also has other properties that are not described by it.
However, this violates assumption 1 and implies either that it is isomorphic to a more complicated mathematical
structure or that it is not mathematical at all.
The latter would be make Karl Popper turn in his grave, since those additional bells
and whistles that make the universe non-mathematical by definition have no observable effects whatsoever.

Another appealing feature of assumption 2 is that it
provides the only answer so far to 
Wheeler's question:
{\it  Why these particular equations, not others?}
Having universes dance to the tune of all possible equations
also resolves the fine-tuning problem of \Sec{TuningSec} once and for all,
even at the fundamental equation level:
although many if not most mathematical structures are likely to be dead and devoid
of SASs, failing to provide the complexity, stability and predictability that SASs require,
we of course expect to find with 100\% probability that we inhabit a
mathematical structure capable of supporting life.
Because of this selection effect, the answer to the question
``what is it that breathes fire into the equations and
makes a universe for them to describe?'' (Hawking 1993)
would then be ``you, the SAS''.

\subsection{What are Level IV parallel universes like?}

The way we use, test and potentially rule out any theory is to
compute probability distributions for our future perceptions
given our past perceptions and to compare these predictions with our observed outcome.
In a multiverse theory, there is typically more than one SAS that 
has experienced a past life identical to yours, so there is no way to determine 
which one is you. To make predictions, you therefore have to compute 
what fractions of them 
will perceive what in the future, which
leads to the following predictions:
\begin{itemize}
\item {\bf Prediction 1:}
The mathematical structure describing our world
is the most generic one that is consistent
with our observations.
\item {\bf Prediction 2:}
Our future observations are the most generic ones that
are consistent with our past observations.
\item {\bf Prediction 3:}
Our past observations are the most generic ones that
are consistent with our existence.
\end{itemize}
We will return to the problem of what ``generic'' means in \Sec{MeasureSec} (the measure problem).
However, one 
striking feature of mathematical structures,
discussed in detail in Tegmark (1997), is that the sort of symmetry and invariance properties
that are responsible for the simplicity and orderliness of our universe tend to be
generic, more the rule than the exception --- mathematical structures 
tend to have them by default, and complicated additional axioms {\etc} 
must be added to make them go away. In other words, because of both this and selection effects, 
we should not necessarily expect life in the Level IV multiverse to be a disordered mess.

\section{Discussion}

We have seen that scientific theories of parallel universes form a four-level hierarchy,
in which universes become progressively more different from ours. They might have
different initial conditions (Level I), different effective physical laws, 
constants and particles (Level II), or different fundamental physical laws (Level IV). 
It is ironic that Level III is the one that
has drawn the most fire in the past decades, because it is the only one that adds no
qualitatively new types of universes. 

Whereas the Level I universes join seemlessly, there are clear demarcations
between those within levels II and III caused by 
inflating space and decoherence, respectively. The level IV universes are completely 
disconnected and need to be considered together only 
for predicting your future, since ``you'' may exist in more than one of them.

\subsection{Future prospects}

There are ample future prospects for testing and perhaps ruling out these multiverse theories.
In the coming decade, dramatically improved cosmological measurements of the 
microwave background radiation, the large-scale matter distribution, \etc, will test Level I 
by further constraining the curvature and topology of space and will test level II by providing 
stringent tests of inflation.
Progress in both astrophysics and high-energy physics should also clarify the extent to which various
physical constants are fine-tuned, thereby weakening or strengthening the case for Level II.
If the current world-wide effort to build quantum computers succeeds, it will provide 
further evidence for Level III, since they would, in essence, be 
exploiting the parallelism of the Level III multiverse for parallel 
computation (Deutsch 1997). 
Conversely, experimental evidence of unitarity violation would rule out Level III.
Finally, success or failure in the grand challenge of modern physics,  
unifying general relativity and quantum field theory, will shed more light on Level IV. 
Either we will eventually find a mathematical structure matching our universe, or we will 
bump up against a limit to  the unreasonable effectiveness of mathematics 
and have to abandon Level IV.

\subsection{The measure problem}
\label{MeasureSec}

There are also interesting theoretical issues to resolve within the multiverse theories,
first and foremost the {\it measure problem}.

As multiverse theories gain credence, the sticky issue of how to compute probabilities in physics
is growing from a minor nuisance into a major embarrassment. If there are indeed many identical
copies of you, the traditional notion of determinism evaporates. You could not compute your own
future even if you had complete knowledge of the entire state of the multiverse, because there is
no way for you to determine which of these copies is you (they all feel they are). All you can
predict, therefore, are probabilities for what you would observe. If an outcome has a probability
of, say, 50 percent, it means that half the observers observe that outcome. 

Unfortunately, it is not an easy task to compute what fraction of the infinitely many observers
perceive what. The answer depends on the order in which you count them. By analogy, the fraction of
the integers that are even is 50 percent if you order them numerically (1, 2, 3, 4, ...) but
approaches 100 percent if you sort them digit by digit, the way your word processor would (1, 10,
100, 1,000, ...). When observers reside in disconnected universes, there is no obviously natural
way in which to order them. Instead  one must sample from the different universes with some
statistical weights referred to by mathematicians as a ``measure''. 

This problem crops up in a mild and treatable manner at Level I, becomes severe at Level II
(see Tegmark 2004 for a detailed review), has
caused much debate at Level III (de Witt 2003, Mukhanov 2005), and is horrendous at Level IV. 
At Level II, for instance, Linde, Vilenkin and others have published predictions for the probability
distributions of various cosmological parameters. They have argued that different parallel
universes that have inflated by different amounts should be given statistical weights proportional
to their volume (\eg, Garriga \& Vilenkin 2001a). 
On the other hand, any mathematician will tell you that $2\times\infty=\infty$, so there is no
objective sense in which an infinite universe that has expanded by a factor of two has gotten
larger. Moreover, a finite universe with the topology of a torus is equivalent to a perfectly
periodic universe with infinite volume, both from the mathematical bird perspective and from the
frog perspective of an observer within it. So why should its infinitely smaller volume give it zero
statistical weight? After all, even in the Level I multiverse, Hubble volumes start repeating
(albeit in a random order, not periodically) after about 10 to the $10^{118}$ meters. 

If you think that is bad, consider the problem of assigning statistical weights to different mathematical
structures at Level IV. The fact that our universe seems relatively simple has led many people to suggest that
the correct measure somehow involves complexity.

\subsection{The pros and cons of parallel universes}

So should you believe in parallel universes?
We have seen that this is not a yes/no question --- rather, 
the most interesting issue is
whether there are 0, 1, 2, 3 or 4 levels of multiverses.
Figure 1 summarizes the evidence that we have discussed 
for the different levels.
The principal arguments against them are that
they are wasteful and that they are weird. 

The wastefulness argument is that multiverse theories
are vulnerable to Occam's razor because they postulate the existence of other worlds that
we can never observe. Why should nature be so wasteful and indulge in such opulence as an
infinity of different worlds? Yet this argument can be turned around to argue for a
multiverse. What precisely would nature be wasting? Certainly not space, mass or
atoms--the uncontroversial Level I multiverse already contains an infinite amount of all
three, so who cares if nature wastes some more? The real issue here is the apparent
reduction in simplicity. A skeptic worries about all the information necessary to specify
all those unseen worlds. 

But an entire ensemble is often much simpler than one of its members. This principle can
be stated more formally using the notion of algorithmic information content. The
algorithmic information content in a number is, roughly speaking, the length of the
shortest computer program that will produce that number as output. For example, consider
the set of all integers. Which is simpler, the whole set or just one number? Naively, you
might think that a single number is simpler, but the entire set can be generated by quite
a trivial computer program, whereas a single number can be hugely long. Therefore, the
whole set is actually simpler. 

Similarly, the set of all solutions to Einstein's field equations is simpler than a
specific solution. The former is described by a few equations, whereas the latter requires
the specification of vast amounts of initial data on some hypersurface. The lesson is that
complexity increases when we restrict our attention to one particular element in an
ensemble, thereby losing the symmetry and simplicity that were inherent in the totality of
all the elements taken together. 

In this sense, the higher-level multiverses are simpler. Going from our universe to the
Level I multiverse eliminates the need to specify initial conditions, upgrading to Level
II eliminates the need to specify physical constants, and the Level IV multiverse
eliminates the need to specify anything at all. The opulence of complexity is all in the
subjective perceptions of observers (Tegmark 1996) --- the frog perspective. From the bird perspective, the
multiverse could hardly be any simpler.

The complaint about weirdness is aesthetic rather than scientific, and it really makes
sense only in the Aristotelian worldview. Yet what did we expect? When we ask a profound
question about the nature of reality, do we not expect an answer that sounds strange?
Evolution provided us with intuition for the everyday physics that had survival value for
our distant ancestors, so whenever we venture beyond the everyday world, we should expect
it to seem bizarre. Thanks to clever inventions, we have glimpsed slightly beyond
the frog perspective of our normal inside view, 
and sure enough, we have encountered bizarre phenomena whenever
departing from human scales in any way: at high speeds (time slows down),
on small scales (quantum particles can be at several places at once), on large scales (black holes),
at low temperatures (liquid Helium can flow upward), 
at high temperatures (colliding particles can change identity), {\etc}.

A common feature of all four multiverse levels is that the simplest and arguably most
elegant theory involves parallel universes by default. To deny the existence of those
universes, one needs to complicate the theory by adding experimentally unsupported
processes and ad hoc postulates: finite space, wave function collapse, ontological
asymmetry, {\etc} Our judgment therefore comes down to which we find more wasteful and inelegant:
many worlds or many words.
Perhaps we will gradually get more used to the weird ways of our cosmos, 
and even find its strangeness to be part of its charm.

\bigskip
{\bf Acknowledgements: }
The author wishes to thank Anthony Aguirre, Aaron Classens, Marius Cohen,
Angelica de Oliveira-Costa, Alan Guth, Shamit Kachru, Andrei Linde, George Musser, 
David Raub, Martin Rees, Harold Shapiro, Alex Vilenkin and Frank Wilczek
for stimulating discussions
and George Musser for extensive editing and improvement of the text of a shorter predecessor 
of this article that was published in {\it Scientific American}. 
This work was supported by
NSF grants AST-0071213 \& AST-0134999,
NASA grants NAG5-9194 \& NAG5-11099,
a fellowship from the David and Lucile Packard Foundation
and a Cottrell Scholarship from Research Corporation.

\section{References} 

\rf\nnn Anglin J R\dualand\nnn Zurek W H;1996;Phys. Rev. D;53;7327

\rfbook;Barbour J B;2001;The End of Time;Oxford Univ. Press;Oxford
 
\rfbook\nnn Barrow J D;1991;Theories of Everything;Ballantine;{New York}

\rfbook\nnn Barrow J D;1992;Pi in the Sky;Clarendon;Oxford
 
\rfbook\nnn Barrow J D\dualand\nnn Tipler F J;1986;The Anthropic Cosmological Principle;Clarendon;Oxford

\rf\nn Bousso R\dualand\nn Polchinski J;2000;JHEP;6;6

\rf\nnn Brundrit G B;1979;Q. J. Royal Astr. Soc.;20;37

\rn\nnn Bucher M A\dualand\nnn Spergel D N 1999, {\it Sci. Am.} {\bf 1/1999}

\rfproc\nn Carter B;1974;IAU Symposium 63;\nn Longair S;Reidel;Dordrecht

\rfbook\nnn Chaitin G J;1987;Algorithmic Information Theory;Cambridge U. P;Cambridge

\rn\nn Cohen M 2003, Master's thesis, Dept. of Philosophy, Ben Gurion University of the Negev, Israel

\rfbook\nnnn Davies P C W;1982;The Accidental Universe;Cambridge U. P;Cambridge

\rfbook\nn Davies P;1993;The Mind of God;Touchstone;{New York}

\rfbook\nn Davies P;1996;Are We Alone?;Basic Books;{New York}

\rfbook\nn Deutsch D;1997;The Fabric of Reality;Allen Lane;{New York}

\rfproc\nn Deutsch D;2003;Science and Ultimate Reality: From Quantum to Cosmos;
\nnn Barrow J D, \nnnn Davies P C W\multiand\nnn Harper C L;Cambridge Univ. Press;Cambridge

\rfproc\nn {de Witt} B;2003;Science and Ultimate Reality: From Quantum to Cosmos;
\nnn Barrow J D, \nnnn Davies P C W\multiand\nnn Harper C L;Cambridge Univ. Press;Cambridge

\rfbook\nn Egan G;1995;Permutation City;Harper;{New York}

\rf\nn Garriga J\dualand\nn Vilenkin A;2001a;Phys. Rev. D;64;023507

\rf\nn Garriga J\dualand\nn Vilenkin A;2001b;Phys. Rev. D;64;043511.

\rfbook\nn Giulini D, \nn Joos E, \nn Kiefer C, \nn Kupsch J,
\nnn Stamatescu I O\multiand\nnn Zeh H D;1996;Decoherence and the Appearance
of a Classical World in Quantum Theory;Berlin;Springer

\rf\nn Guth A\dualand\nnn Steinhardt P J;1984;Sci. Am.;250;116

\rfbook\nn Hawking S;1993;A Brief History of Time;Touchstone;{New York}

\rfbook\nn Lewis D;1986;On the Plurality of Worlds;Blackwell;Oxford

\rf\nn Linde A;1994;Sci. Am.;271;32

\rfproc\nn Maldacena J;2003;Science and Ultimate Reality: From Quantum to Cosmos;
\nnn Barrow J D, \nnnn Davies P C W\multiand\nnn Harper C L;Cambridge Univ. Press;Cambridge

\rfbook\nn Nozick R;1981;Philosophical Explanations;Harvard Univ. Press;Cambridge

\rfbook\nnn Rees M J;2002;Our Cosmic Habitat;Princeton Univ. Press;Princeton

\rfbook\nn Rucker R;1982;Infinity and the Mind;Birkhauser;Boston

\rn\nn Schmidthuber J 1997, {\it A Computer Scientist's View of
Life, the Universe, and Everything}, in 
{\it Foundations of Computer Science: Potential-Theory-Cognition, 
Lecture Notes in Computer Science}, C. Freksa, ed., 
(Springer: Berlin), http://www.idsia.ch/~juergen/everything/html.html

\rfbook\nn Smolin L;1997;The Life of the Cosmos;Oxford Univ. Press;Oxford

\rf\nnn Steinhardt P J\dualand\nn Turok N;2002;Science;296;1436

\rfprep\nn Susskind L;2003;hep-th/0302219 

\rf\nn Tegmark M;1996;Found. Phys. Lett.;9;25

\rf\nn Tegmark M;1997;Class. Quant. Grav.;14;L69

\rf\nn Tegmark M;1998;Ann. Phys.;270;1

\rf\nn Tegmark M;2000;Phys. Rev. E;61;4194 

\rf\nn Tegmark M;2002;Science;296;1427

\rf\nn Tegmark M\dualand\nnn Wheeler J A;2001;Sci.Am.;2/2001;{68-75} 

\rfprep\nn Tegmark M;2004;astro-ph/0410281

\rfbook\nnn Wigner E P;1967;Symmetries and Reflections;MIT Press;Cambridge

\rfbook\nnn Wolfram S;2002;A New Kind of Science;Wolfram Media;{New York}

\rf\nnn Zeh H D;1970;Found. Phys.;1;69

\rfproc\nn Zurek W;2003;Science and Ultimate Reality: From Quantum to Cosmos;
\nnn Barrow J D, \nnnn Davies P C W\multiand\nnn Harper C L;Cambridge Univ. Press;Cambridge

\end{document}